\documentclass[twocolumn,showpacs,preprintnumbers,aps,prstab]{revtex4}
\usepackage{graphicx}
\usepackage{amsmath}
\usepackage{color}

\begin{document}
\title{ 
Coupling of Laser-Generated Electrons with Conventional Accelerator Devices
}

\author{P. Antici} 
\author{A. Bacci}
\author{E. Chiadroni} 
\author{M. Ferrario} 
\author{A.R. Rossi} 
\email{andrea.rossi@lnf.infn.it}
\affiliation{INFN-LNF, Frascati, Roma, Italy}
\author{C. Benedetti} 
\altaffiliation[Now at]{ LBNL, Berkeley, CA 94720-8211, USA}
\affiliation{University of Bologna \& INFN, Bologna, Italy}
\author{L. Lancia} 
\author{M. Migliorati} 
\author{A. Mostacci} 
\author{L. Palumbo}
 \affiliation{SAPIENZA University of Rome, Roma, Italy}
\author{L. Serafini} 
\affiliation{INFN-MI, Milano, Italy}
%
\begin{abstract}
Laser-based electron acceleration is attracting strong interest from the conventional accelerator community due to its outstanding characteristics in terms of high initial energy, low emittance and high beam current. Unfortunately, such beams are currently not comparable to those of conventional accelerators, limiting their use for the manifold applications that a traditional accelerator can have. Besides working on the plasma source itself, a promising approach to shape the laser-generated beams is coupling them with conventional accelerator elements in order to benefit from both, a versatile electron source and a controllable beam. 
 
In this paper we show that some parameters commonly used by the particle accelerator community must be reconsidered when dealing with laser-plasma beams. Starting from the parameters of laser-generated electrons which can be obtained nowadays by conventional multi hundred TW laser systems, we compare different conventional magnetic lattices able to capture and transport those GeV electrons. From this comparison we highlight some important limit of the state-of-the-art plasma generated electrons with respect to conventional accelerator ones. Eventually we discuss an application of such beams in undulators for Free Electron Lasers (FELs), which is one of the most demanding applications in terms of beam quality.
\end{abstract}
\maketitle

\section{INTRODUCTION}
\label{sec:intro}

Particle accelerators have been one of the most important scientific instruments for the research activity of the last 50 years. Ever since the beginnings, an exponential increase of the beam energies has accompanied the technological progress of the accelerators. However this tendency seems now destined to reach saturation \cite{uno}. In fact, the acceleration of charged particles relies on the use of radio-frequency longitudinal electric fields. They are generally produced in cavity resonators and their amplitude is proportional to the energy gain of the charged particles. However, if the electric fields in the cavities reach some tens of MV/m, a breakdown of the material, producing a leak of electrons from the metal surface, will damage the resonators, preventing the particle acceleration. As a consequence, the energy gain per unit length in conventional accelerators is limited, and high energies can only be reached by additional acceleration modules, thus explaining the large dimensions of some state-of-the-art accelerators.

A completely new approach toward particle acceleration, which became feasible only in recent years, is based on the laser-plasma interaction \cite{Esa09}. In comparison with conventional accelerators, a plasma can generate and support electric fields up to TeV/m, thus giving the possibility to accelerate a particle beam to energies up to few GeV in a distance of few cm instead of a hundred meters. Several experiments up to now have shown the capabilities of laser-plasma acceleration, producing particle beams of high quality and mid to high energy \cite{Ged04,Lee06}. Currently, laser-plasma accelerators can probuce beams of 1 GeV over 3 cm of acceleration length \cite{Lee06}. These laser-produced particle beams possess a number of outstanding properties, such as ultra-short pulse duration, high peak currents and excellent emittance at the plasma-vacuum interface. Given these unique beam properties, and the compactness of the acceleration scheme that will results in a strong advantage in terms of size and cost of the global accelerating infrastructure, the field of laser-based particle acceleration has attracted much attention.

In this paper we present a design of a particle transport system based on conventional magnetic devices (solenoids and quadrupoles) that is able to capture and transport such novel beams. Moreover, we will show the present limitations of these beams, indicating the way to improve their quality. In fact, while currently achieved parameters do not allow for a realistic conceptual study yet, we find that our simulation studies can already give a useful guidance. 

In particular, Sec.~\ref{sec:1} reminds the peculiarities of laser generated electron by self-injection mechanism and summarizes the main characteristics of the beam used as the starting point for all the simulation studies. Section~\ref{emittance} presents a consistent definition of normalized emittance for beams with high energy spread and high divergence, typical not only of laser plasma sources, but also of wakefields accelerators. Conventional transport lines for such beams are critically discussed in Sec.~\ref{sec:quads}-\ref{sec:cut}, while Sec.~\ref{sec:FEL} shows the weak points of using those beams in conventional undulators for the generation of FEL radiation.

\section{LASER-GENERATED ELECTRON BEAMS }
\label{sec:1}

There are many physical processes in laser-plasma interaction that can be exploited to produce high energy beams. In what follows, we will focus on electron produced in the bubble regime \cite{Gor05,Lu07} which occurs when the laser waist $w_0$ is smaller than the plasma wavelength $\lambda_p$ and the laser parameter $a_0$ sufficiently high (e.g. $a_0\approx$~2.5-3). Nevertheless our conclusions will be valid also for most other experimental setups, such as Laser Wake Field Acceleration LWFA in a mildly non linear regime or plasma wakefield accelerators. 

When a high power laser impinges on an under dense plasma, it propagates inside it and, by ponderomotive force, creates a volume almost completely depleted of electrons (i.e. the bubble). Due to the presence of the background positive ions (assumed to remain still), strong electric fields are present, whose peak longitudinal value scales as 
\begin{equation} \label{maxfield}
E_z[{\rm V/m}]\approx 100\sqrt{n_0[{\rm cm^{-3}}\,] a_0}\,,
\end{equation}
where $n_0$  is the plasma density and $a_0$  the laser parameter. Such a scaling law  assumes that the bubble is stable and matched and it requires the bubble to be roughly spherical. Thus the peak value for the longitudinal force (responsible for the electrons acceleration) has the same order of magnitude of its transverse component that gives the focusing strength for a relativistic electron \cite{Esa09}. Therefore the resulting focusing is much more intense than in any conventional accelerator.  Some electrons are injected into the strongly accelerating longitudinal field at the bubble bottom: the details of injection depend on different physical processes and are thus hard to control experimentally. In some regimes, injection will occur only once for a brief time \cite{Bru10,Sea10}, in some others, it will be subject to a 
non uniform trend, as the one considered hereafter, while in others it would happen  continuously. There are also some experimental schemes where injection can be controlled by a counter propagating laser \cite{Rec09}, at the price of a very low injected charge. In a recent paper, injection is started by tailoring the plasma longitudinal density \cite{Gon11}. 

If the plasma is assumed to be cold, the transverse average normalized electrons momentum at the injection into the bubble can reach a magnitude of order $a_0$. Only electrons with a longitudinal component of the velocity greater than the bubble speed will be captured and accelerated; therefore plasma electrons injected in the bubble are quite hot. 

Such electrons posses, as well, a transverse (thermal) momentum of the same order of magnitude of the longitudinal one, that is few MeV/c in our numerical simulations. Typical plasma accelerated beams (considered in this paper) reach energy of the order of the GeV and therefore their divergence approaches few mrads when they exit from plasma.
On the other hand, typical electrons produced in a RF photoinjector posses a transverse momentum of few eV/c when extracted from the photocathode; the subsequent acceleration brings the beam to a few MeV energy and the divergence down to $10^{-3}$ mrad.

The high beam temperature and the lack of divergence reduction deeply influence the behavior of the beam as soon as it leaves the plasma channel. It is then clear also why increasing the energy achieved by electrons in the plasma would have beneficial effects on the average beam divergence as well.

To summarize, the electron relativistic momentum at injection into the bubble is responsible of a high divergence and a large uncorrelated energy spread, as compared to bunches in conventional accelerators. Moreover, since the injection could occur along the whole plasma channel (typically few millimeters), we can also expect a large correlated energy spread, due to the very high field gradient.

The presence of a strong focusing field implies that the beam is emittance dominated; its transverse distribution, both in charge density and in momentum, is expected to be well represented by a Gaussian. From the transverse envelope equation, after the plasma-vacuum interface, we can expect that the beam transverse size would increase, due to free diffraction, as \cite{r1}:
\begin{equation} \label{sigma_s}
\sigma_x(s)=\sqrt{\sigma_0^2+2\sigma_0 \sigma'_{0}s+\left( \frac{\varepsilon^2}{\sigma_0^2}+{\sigma'_{0}}^2 \right)s^2}\,,
\end{equation}
where $s$ is the longitudinal coordinate, $\varepsilon$ the geometric r.m.s. emittance, $\sigma_0 \equiv \sigma_x(0)$ the initial size and $\sigma'_0 \equiv d \sigma_x/ds (0)$ its first derivative.      

In order to design a transport system for laser-plasma electron beams, we will focus on the capture and transport of electrons generated by 100 TW class lasers, which are currently establishing worldwide in different labs, being also a test facility for larger scale facilities currently in planning phase, such as 10 PW lasers. The global parameters of the obtained bunch are reported in Table \ref{tab1}, which refers to the following structure: a 200 TW laser with wavelength $\lambda =0.8$~$\mu$m, contrast ratio of 10$^{10}$, pulse duration (FWHM) $\tau =30$ fs, and waist=15.5 $\mu$m, delivering an intensity of $I=5\times 10^{19}$ W/cm$^2$, similar to what we would expect to obtain with the FLAME laser, currently under commissioning at the Frascati National Laboratories \cite{flame}. For the gas jet, we have considered the gas jet installed at FLAME, which produces a plasma of length 4.1 mm and electron density $3 \times 10^{18}$ cm$^{-3}$. These parameters where chosen to produce the highest beam energy with the available laser power \cite{Ben09}.
%
\begin{table}[!h]
\centerline{
\begin{tabular}{||c|c||}
\hline\hline
Charge & 700 pC \\ 
\hline
Energy & 910 MeV \\
\hline
Energy spread $\sigma_{\varepsilon}$ & 6.4 \% \\
\hline
Bunch length & 2 $\mu$m\\
\hline
Transverse size $\sigma_{x}$ & 0.5 $\mu$m \\
\hline
Transverse divergence $\sigma_{x'}$ & 3 mrad \\
\hline
Emittance $\varepsilon$ & 1.4$\times10^{-3}$ mm mrad\\
\hline
Normalized emittance $\varepsilon_n$ & 2.5 mm mrad \\
\hline\hline
\end{tabular}}
\caption{ Main beam parameters used for the design of the electron beam transport. The beam is assumed of circular spot and only the $x$ transverse coordinate is considered.}
\label{tab1}
\end{table}

\begin{figure}[!h]
\centerline{\includegraphics[width=7.5cm]{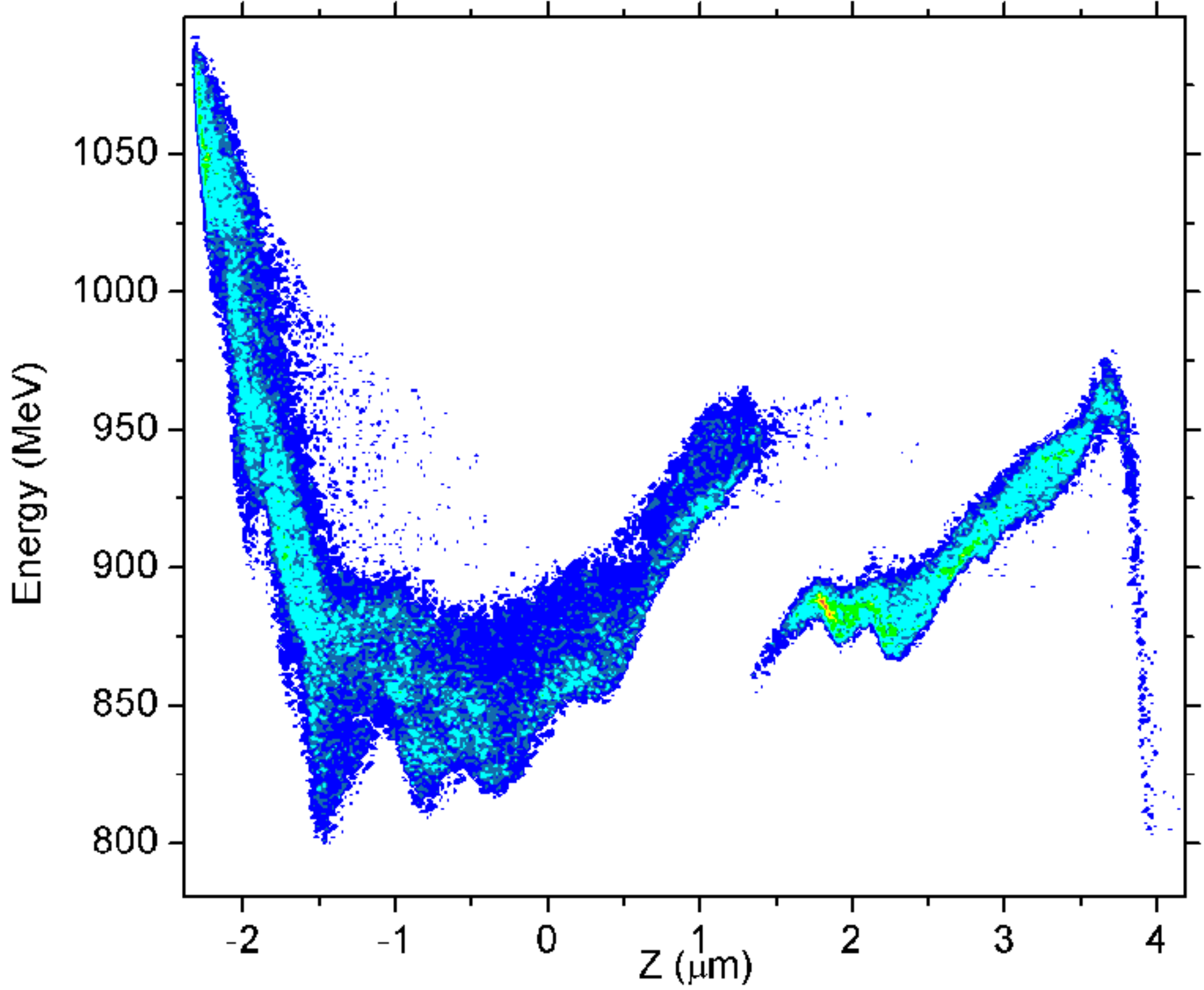}} 
\caption{(Color online) Longitudinal phase space distribution obtained with the PIC code ALaDyn and used for the transport system design.}
\label{fig1}
\end{figure}

Regarding the beam distribution at the exit of the plasma (i.e. at the entrance of the transport channel), we have considered the one given by the 3D - PIC code ALaDyn \cite{aladyn}. With the above mentioned  laser and plasma characteristics, the longitudinal phase space bunch distribution is shown in Fig.~\ref{fig1}. Concerning the transverse phase space, the beam distribution appears almost Gaussian in the transverse position and divergence, as expected for an emittance dominated beam; at the plasma exit, the beam is in a waist and its position and divergence are therefore uncorrelated.
%

\section{HIGH DIVERGENCE AND HIGH ENERGY SPREAD BUNCHES }
\label{emittance}

In order to obtain a beam that can easily be transported in a conventional FODO line, we have to match the initial bunch parameters at the exit of the plasma with those required by the FODO cell. 

\begin{figure}[!ht]
\centerline{\includegraphics[width=7.5cm]{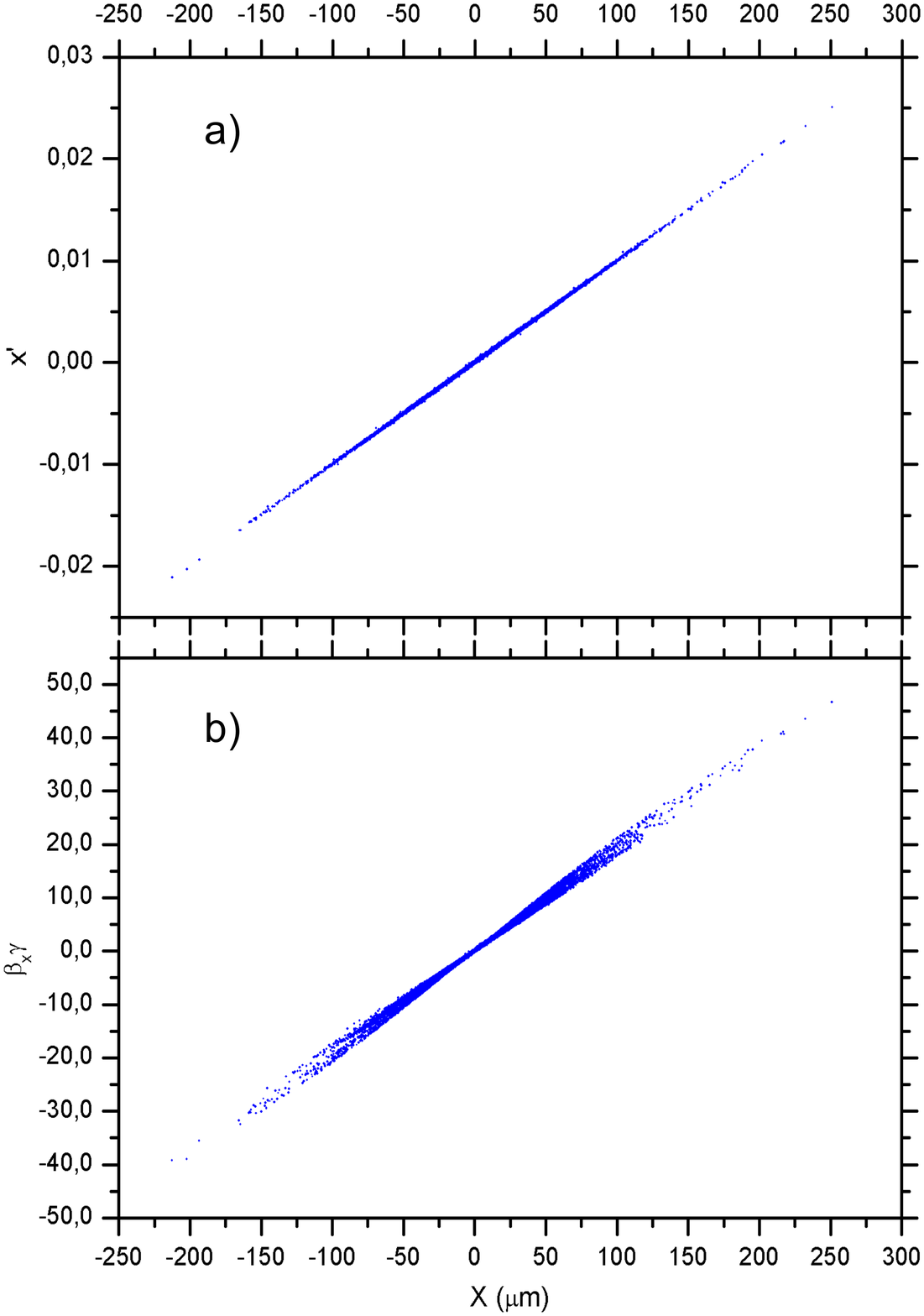}} 
\caption{(Color online) Transverse phase space distribution after a drift of 1 cm obtained with TSTEP.}
\label{fig4}
\end{figure}

Once the beam comes out from the laser-plasma interaction region, it suddenly passes from the extremely intense focusing fields inside the plasma to free space. A general strategy to control the beam should be that of using strong magnetic fields by means of quadrupoles or solenoids, as close as possible to the interaction point. However, from the parameters of Table \ref{tab1}, the Twiss beta function, $\beta_T = \sigma_x^2/\varepsilon$ is as low as a fraction of $\mu$m, many order of magnitude lower than what is generally reachable in conventional accelerators. 
Therefore, even if the initial transverse beam correlation is zero, it increases very rapidly in a drift. Figure \ref{fig4}a shows the transverse phase space after a 1 cm drift. The figure has been obtained by using the well established macroparticle code TSTEP \cite{tstep}, a derivative of PARMELA \cite{parmela}. The transverse dimension at this point is about 60 times higher than the initial value and it keeps increasing at a constant rate; conventional magnets have not enough strength to counteract such a behavior and the beam control is difficult. On top of that, at least a short drift has necessarily to be foreseen before using magnetic quadrupoles or solenoids for beam matching, because of the physical dimensions of the devices at the interaction region.

\begin{figure}[!ht]
\centerline{\includegraphics[width=7.5cm]{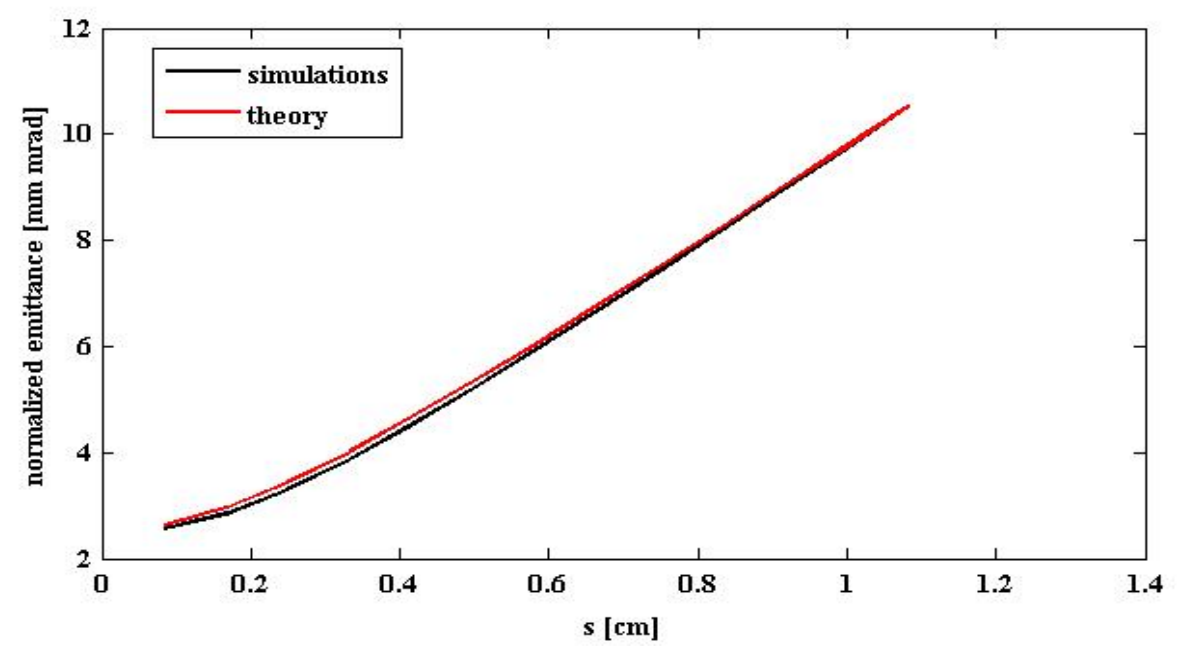}} 
\caption{(Color online) Increase of the transverse normalized emittance along a drift of 1 cm: theory and simulation results.}
\label{fig5}
\end{figure}

The strong increase of the transverse beam size and the difficulty to control the Twiss parameters are not the only problems in designing a matching line. One apparently strange behavior of the bunch that turned out from macroparticle simulations is shown in Fig. \ref{fig5}, where the black curve displays the transverse normalized emittance in the 1 cm drift following the interaction region (the choice of 1 cm has been done just as a reference for the analysis of a general behavior of the bunch in the drift immediately following the interaction point).

If space charge and wake field effects are neglected, as in this case, one should expect a constant emittance in a drift. This is true for the emittance $\varepsilon$, since the transverse motion is uncoupled with others planes, and it  is generally thought to be true also for the normalized emittance $\varepsilon_n$, since it is usually obtained, in conventional accelerators, as $\varepsilon_n=<\gamma> \varepsilon$ with $<\gamma>$ the {\em average} relativistic $\gamma$ factor. This way of evaluating the normalized emittance has been used for laser-plasma generated electron beams in international journal papers (see e.g. \cite{Bru10,Sea10,emit}). 
However, a more detailed analysis of the definition of the normalized emittance shows that more care has to be taken when evaluating this quantity and that the bunches coming out from the laser-plasma interaction cannot be treated as conventional beams.

In order to calculate the right expression for the normalized emittance, let us start from its definition:
\begin{equation}
\varepsilon_n^2 = <x^2> <\beta^2 \gamma^2 x'^2>-<x \beta \gamma x'>^2
\end{equation}
with $\beta$ and $\gamma$ the particle relativistic factors, and $x$ and $x'$ the transverse position and its divergence. If the correlation between the energy and transverse position is negligible, as in a drift without collective effects, we can write
\begin{eqnarray}
\label{eq2}
\varepsilon_n^2 &=& <\beta^2 \gamma^2> <x^2> < x'^2>+ \\ \nonumber
&&- < \beta \gamma>^2<x x'>^2
\end{eqnarray}
The definition of relative energy spread $\sigma_{E}$ allows us to write
\begin{equation}
\sigma_{E}^2 = \frac{<\beta^2 \gamma^2>-< \beta \gamma>^2} {<\gamma>^2},
\end{equation}
which can be inserted in Eq.~(\ref{eq2}) to give
\begin{eqnarray}
\varepsilon_n^2 &=& <\gamma>^2  \sigma_{E}^2<x^2> < x'^2>+ \\ \nonumber
&&+ < \beta \gamma>^2
(<x^2><x'^2>-<x x'>^2).
\end{eqnarray}
Assuming relativistic electrons $(\beta=1)$, we get
\begin{equation}
\varepsilon_n^2 = <\gamma>^2 \left( \sigma_{E}^2 \sigma_x^2 \sigma_{x'}^2 + \varepsilon^2 \right).
\label{emittn}
\end{equation}
If the first term on the right hand side is negligible, then the normalized emittance is the usual $<\gamma> \varepsilon$ value. For a conventional accelerator this is generally true: if we consider for example the bunch of the SPARC photoinjector \cite{sparc} we get that, at low energies (5 MeV), the first term gives a contribution to the normalized emittance a factor 10$^3$ lower than the emittance of the second term, and at higher energies this factor is even higher: at about 150 MeV it is in the order of 10$^5$.

If we consider the laser-plasma beam of Table \ref{tab1}, we find that the first term on the right hand side of Eq.~(\ref{emittn}), at the plasma-vacuum interface, has the same order of magnitude as that of conventional accelerators at low energies, i.e. about a factor $10^3$ lower than the geometric emittance; anyway, after only 1 cm of drift, due to the rapid increase of the bunch size, the first term becomes predominant, being more than three times the second one. Therefore the normalized emittance increases in a drift. 

At the plasma-vacuum interface, the beam is in a waist, thus one can write in Eq.~(\ref{sigma_s})  $\sigma'_{0}$=0 and $\varepsilon^2/\sigma_0^2=\sigma_{x'}^2$. In the limit $s\gg1$, Eq.~(\ref{sigma_s}) reads $\sigma_x(s) \approx \sigma_{x'} s$ and 
Eq.~(\ref{emittn}) becomes
\begin{equation}
\varepsilon_n^2 \approx <\gamma>^2 \left( \sigma_{E}^2  {\sigma_{x'}^4} s^2 + \varepsilon^2 \right).
\end{equation}

This last equation is represented in Fig. \ref{fig5} by the red curve and it shows a very good agreement with the results of the multiparticle simulation code. Notice that for protons and heavier ionic species, the assumption of relativistic motion done in deriving Eq.~(\ref{emittn}) is usually not valid; this worsen the effect on normalized emittance and produces a lengthening of the produced bunch.
\begin{figure}[!h]
\centerline{\includegraphics[width=7.5cm]{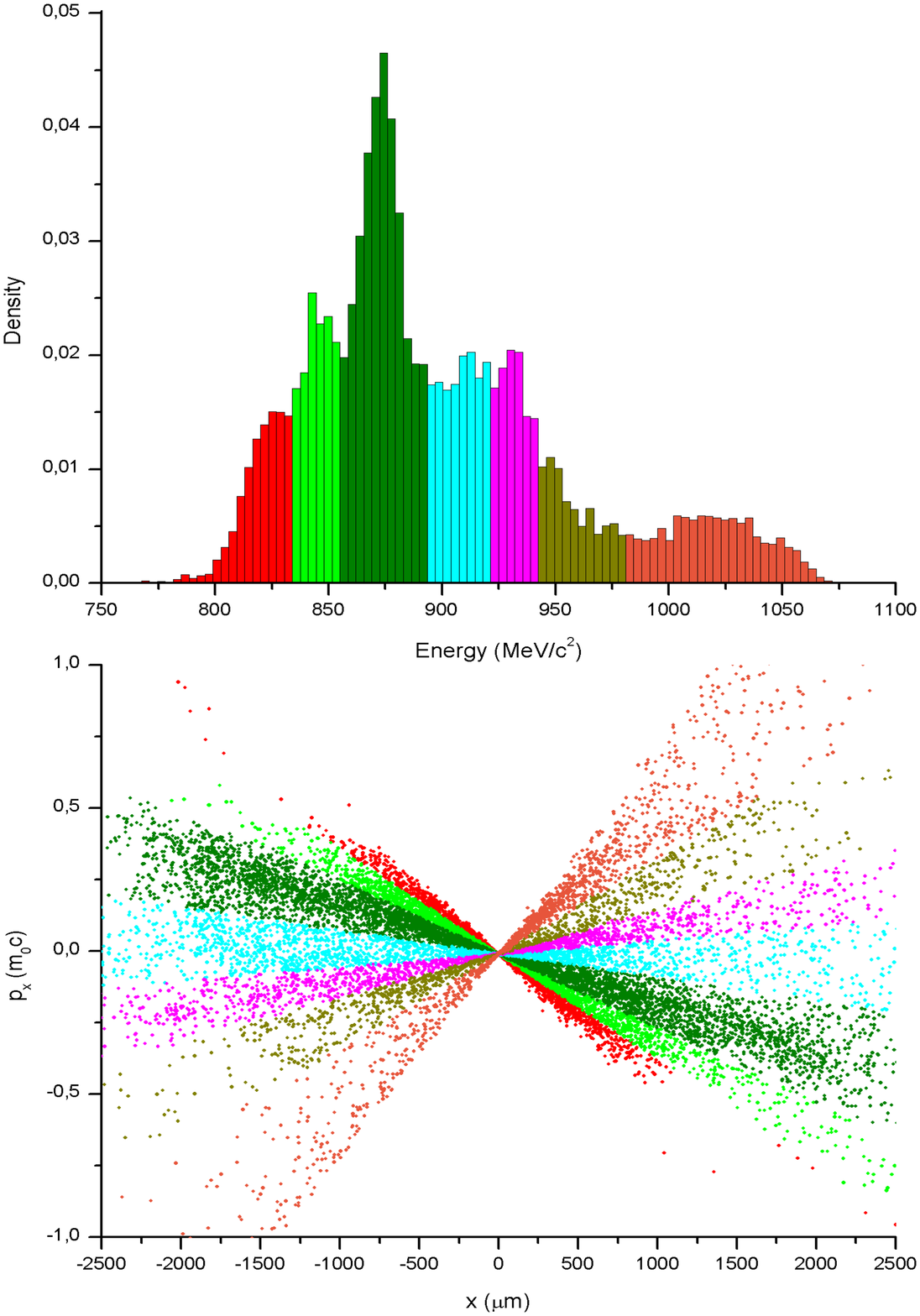}} 
\caption{(Color online) Bunch spectrum (up) and de-correlated transverse phase space (down). Here the free drift is about 1 m long, in order to enhance the effects of chromatism.}
\label{fig4a}
\end{figure}

The physical reason behind such a dramatic emittance dilution is readily understood when we realize that the betatron frequency of a beam critically depends on its energy. During the drift, each chromatic component rotates with its own velocity in the transverse phase space, spreading out the area occupied by the whole bunch, as is clearly shown by Fig. \ref{fig4}b; the resulting {\em projected} normalized emittance then becomes a function of both position and energy spread. In Fig. \ref{fig4a}a we can see a spectrum of the bunch where a few (arbitrarily chosen) energetic components have been outlined using different colors: the chromatic effect is then pictorially shown in Fig \ref{fig4a}b where the transverse phase space is reproduced after removing the $x-p_x$ correlation. Notice that, if the divergence $x'$ is defined as $x'\approx p_x/<p>$ (see, for example, \cite{r2}), such an emittance dilution is {\em not} detected, as seen in Fig. \ref{fig4}a.

\begin{figure}[!ht]
\centerline{\includegraphics[width=7.5cm]{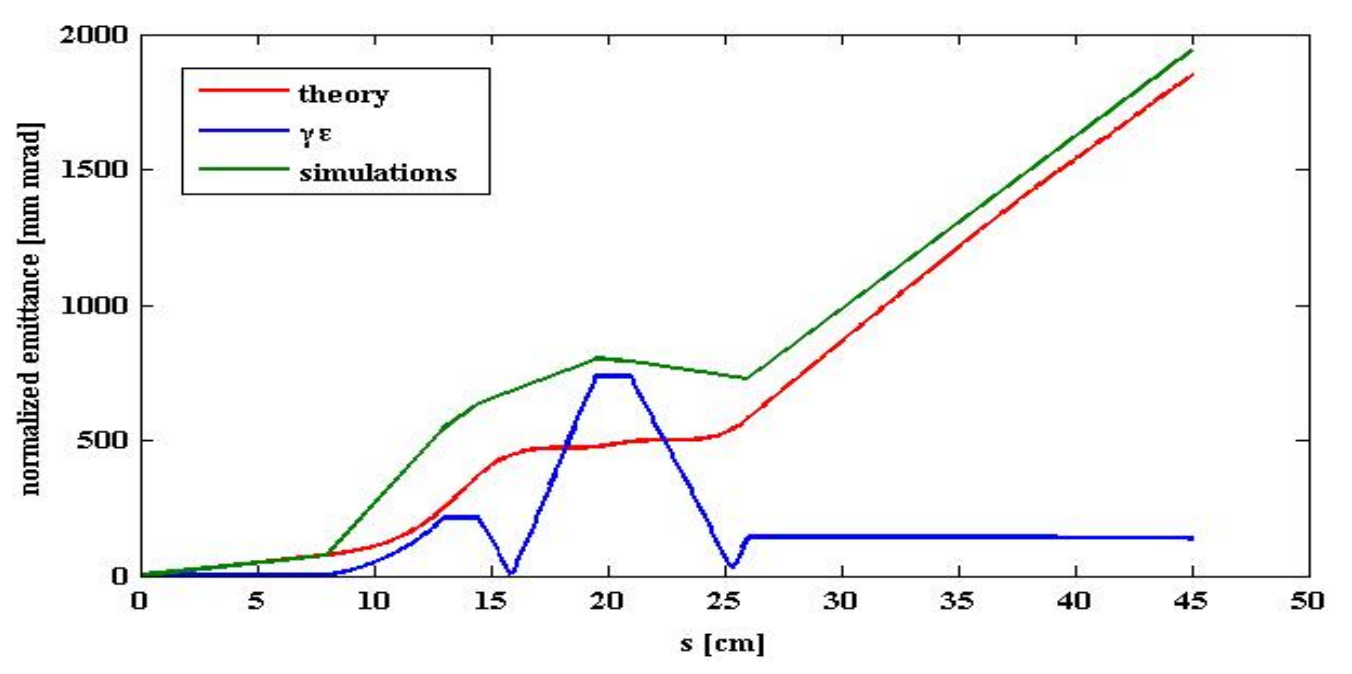}} 
\caption{(Color online) Normalized emittance along a quadrupole magnetic transport line: comparison between simulations and theory.}
\label{fig7}
\end{figure}

Equation~(\ref{emittn}) can be also used to predict the normalized emittance behavior in presence of magnetic fields, for examples quadrupoles, when there is some coupling between energy and transverse coordinate. In Fig. \ref{fig7} we show a comparison between Eq.~(\ref{emittn}), the expression $<\gamma> \varepsilon$ and  the normalized emittance obtained with the macroparticle simulation code. The agreement between theory and simulation is worse than in the case of a simple drift; nevertheless Eq.~(\ref{emittn}) remains a valid analytical tool to quickly determine the behavior of the normalized emittance in a transport line without running macroparticle tracking codes.
   
\begin{figure}[!h]
\centerline{\includegraphics[width=7.5cm]{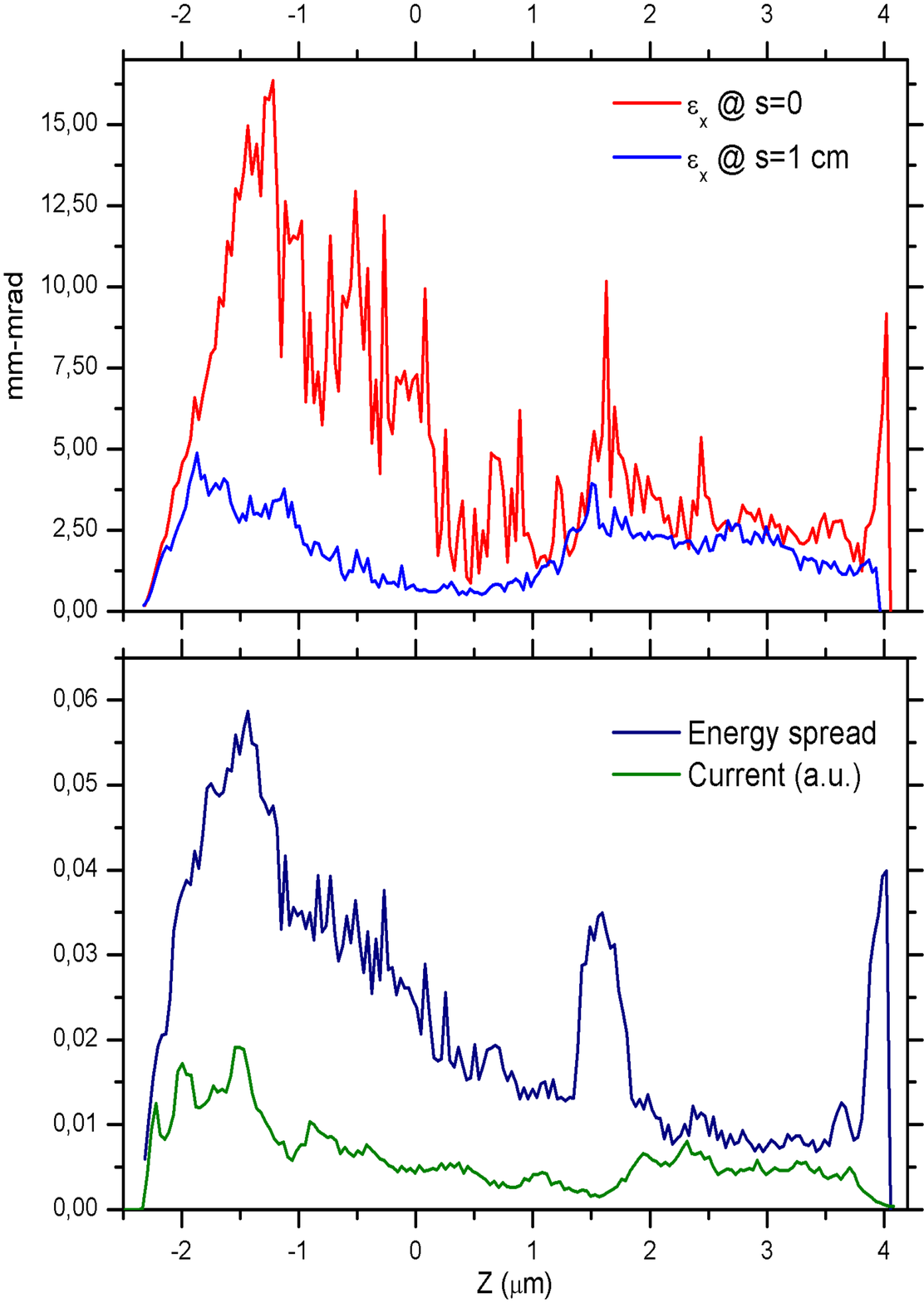}} 
\caption{(Color online) Comparison of slice emittance before and after free propagation (upper plot) together with slice energy spread (lower plot). The green curve reports the slice current in order to show the longitudinal charge distribution, The number of slices is arbitrarily set to 200.}
\label{fig_emislice}
\end{figure}

We must underline that chromatic effects could be different for slice normalized emittance, which is the critical parameter for some applications; for example in FELs the longitudinal slices are of the order of the cooperation length. If such slices are sufficiently thin, the contribution of the correlated energy spread to the slice emittance is usually negligible with respect to uncorrelated energy spread (which depends on thermal emittance). Therefore, concerning the slice parameters, laser-plasma beams may not differ much from beams from conventional accelerators.
In the upper plot of Fig. \ref{fig_emislice}, a comparison between slice emittance at $s=0$ and $s=1$ cm is shown, while in the lower plot we show the slice energy spread. It is readily apparent how slice emittance is almost conserved in the longitudinal range between 2 and 4 $\mu$m, i.e. where the slice energy spread is relatively low ($\le 1\%$). 

The results here presented demonstrate that the laser-plasma electron beams have a behavior quite different from beams of conventional accelerators. In a drift immediately following the laser-plasma interaction point, we find the uncommon situation in which the normalized emittance not only increases, but with a rate as high as of 1000 mm mrad/m. This is a rather new result that must be carefully considered in the design of a beam transport system. It depends on the combination of the RMS values of energy spread, beam divergence and beam size. To match such a beam to a transport line, without spoiling the bunch normalized emittance, requires the optical elements to have a focal length of the order of the bunch $\beta$ function, which is impossible to achieve with conventional technology. Moreover, since the bunch high chromaticity,  any possible viable matching would be valid only for that  portion of charge possessing the proper energy: in fact, an achromatic focusing device is not realizable by magnetic quadrupoles or solenoids \cite{Cou72}. 

In order to overcome these problems, it is necessary to act on the plasma-laser source itself, because conventional magnetic transport line are inadequate, as we will see in the next sections. Plasma lenses could be employed as well~\cite{Tom_ip}.

\section{QUADRUPOLE MATCHING LINE}
\label{sec:quads}
We have investigated two ways for controlling and matching the beam to a FODO transport line: by using quadrupoles or solenoids. A quadrupole matching line is more commonly used in conventional particle accelerators due to the lower cost as compared to a solenoid line. For the design of our matching system, we have first used the code TRACE 3-D \cite{trace}, an interactive beam dynamics program that calculates the envelopes of a bunched beam, including linear space charge forces, through a user defined transport system. Once the beam line has been optimized, the magnet parameters have been inserted in the tracking code TSTEP to account for the beam distribution function.

\begin{figure}[!h]
\centerline{\includegraphics[width=7.5cm]{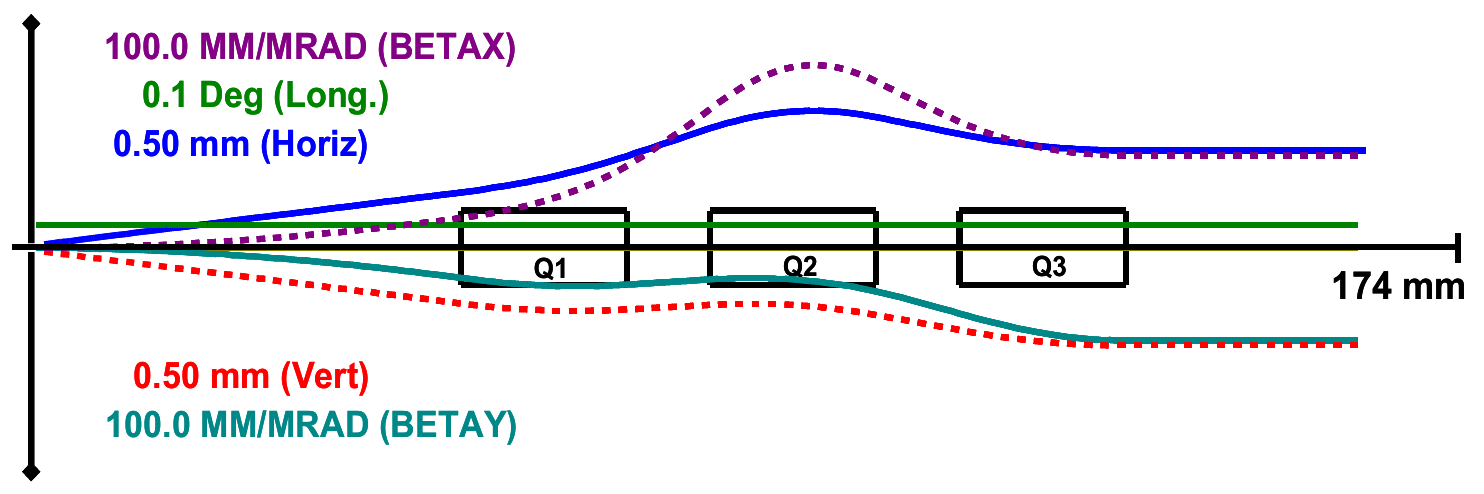}} 
\caption{(Color online) Matching transport line with quadrupoles given by TRACE 3-D. Axis limits for the different shown parameters are indicated by the normalized ordinate in figure.}
\label{fig8}
\end{figure}

In Fig. \ref{fig8} we show the optimized beam line obtained with TRACE 3D. Starting from very low Twiss $\beta_T$ values, of the order of 0.2 mm, by using a matching system of three quadrupoles, with a total length of 174 mm, we manage to obtain at the end of the transport line a bunch with Twiss parameters $\alpha_T$=0 and $\beta_T$=45 m. Such final beam parameters can be transported with a periodic FODO system without particular problems. Also the transverse dimensions of the bunch allow a conventional transport after the matching line.

For a deeper investigation of the beam dynamics, we have introduced the magnets parameters and dimensions into the code TSTEP. The bunch remains confined below 0.4 mm and the behavior agrees with the results of TRACE 3D.

Even if the quadrupole matching line is quite attractive, because of its simplicity and cheapness, two problems arise by using this kind of transport system: one is a pure optical problem, the other is related to the definition of normalized emittance.

From the point of view of the optics, the strength of the quadrupole gradient necessary to contain the divergence of such beams reaches values as high as 5000 T/m, a factor 10 higher than the maximum gradient currently available and obtained using permanent magnet quadrupoles \cite{pmq}. We tried to reduce the quadrupole gradient to values of 500 T/m, that is the actual state of the art, and increase the number of quadrupoles, but we did not manage to control the beam explosion. 

\begin{figure}[!th]
\centerline{\includegraphics[width=7.5cm]{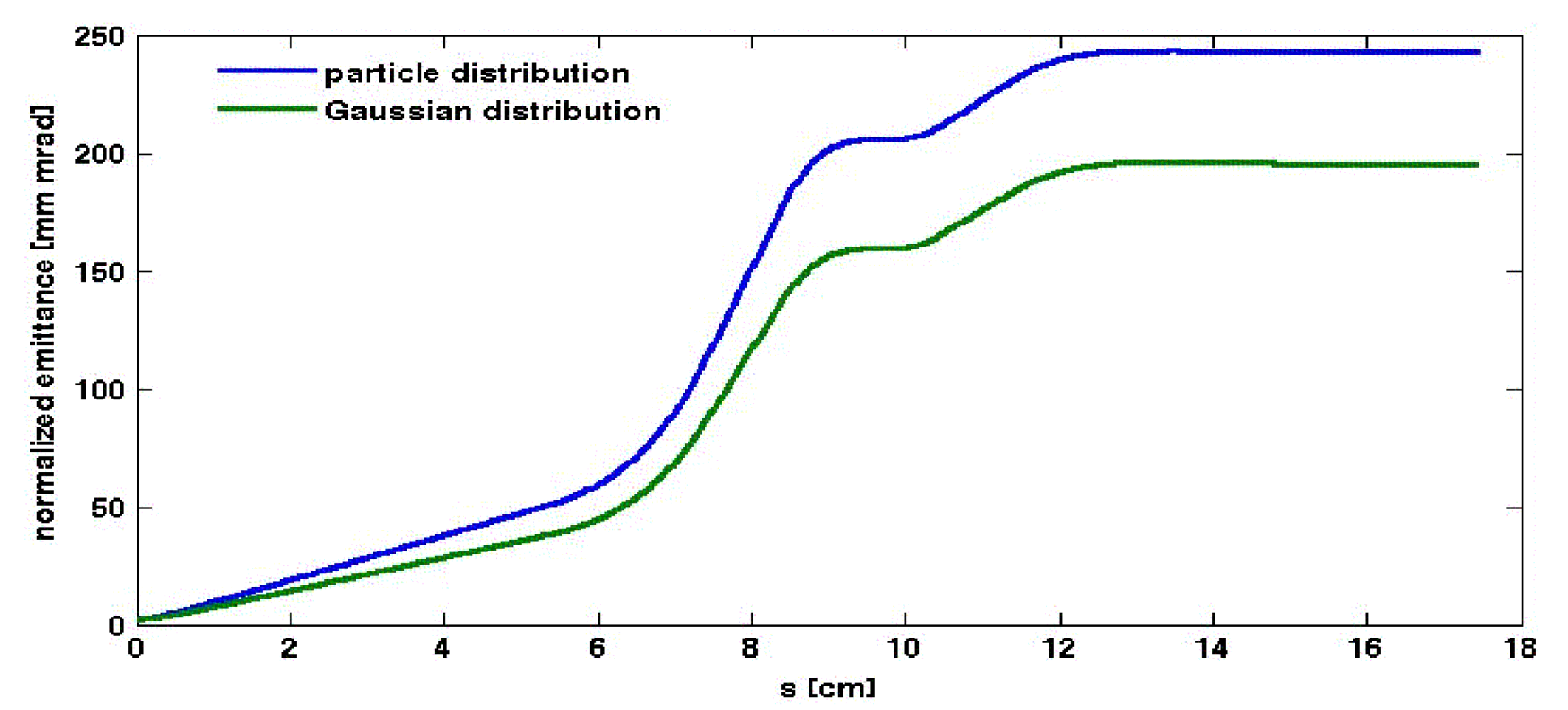}} 
\caption{(Color online) Transverse normalized emittance along the quadrupole line obtained with TSTEP.}
\label{fig9}
\end{figure}

The second problem is related to the normalized emittance $\varepsilon_n$. Using the quadrupole matching line that has been optimized for the optics, even if we allow the presence of permanent magnet quadrupoles with gradients of 5000 T/m, we obtain that the normalized emittance at the exit of the line reaches values up to 250 mm mrad, excessively high for an accelerator. In Fig. \ref{fig9} we show the results of the simulations obtained with TSTEP starting from the distribution function of Fig. \ref{fig1} (blue line) and with an ideal Gaussian distribution (green line). As expected, the main problem is not related to the kind of distribution, but to the initial beam parameters.

\section{SOLENOID MATCHING LINE}
\label{sec:solenoid}

An alternative to quadrupoles is the use of solenoids that, acting over a longer distance than quadrupoles, allow a smoother control of the bunch transverse size. As in the previous section, we have first optimized the beam line by using TRACE 3-D. A solenoid of about 20 cm, located very close to the laser-plasma interaction point, is able to control the beam size and the Twiss beta function. At the exit of the solenoid we obtain a beta function of about 88 m in both directions, and a transversely uncorrelated bunch ($\alpha_T=0$).

We have then performed simulations with TSTEP, obtaining a transverse size behavior similar to that  of quadrupoles, remaining confined below 0.4 mm.

A transport line with solenoids is more expensive and complicated with respect to one with quadrupoles. However, in this case, the intensity of magnetic field inside the solenoid, even though extremely high, seems feasible with the actual state of the art technology. In fact, the optimization with TRACE 3-D requires a longitudinal magnetic field of about 45-50 T, an unusual value for conventional accelerators, since e.g. for SPARC \cite{sparc1} and LCLS \cite{lcls} photoinjectors it is below 0.5 T. Such high field solenoids have been under investigation by using high temperature superconductors, and a conceptual design of a solenoid of 45 T  has been recently presented \cite{50t}. However, as expected, the normalized beam emittance endures the same effect of uncontrollable increase as the one shown in Fig. \ref{fig9}. At the end of the beamline it reaches a value of about 200 mm mrad, not much different from the result with quadrupoles.

Even though a transport system by using a strong solenoidal field seems feasible from the optics point of view, the problem of an uncontrollable increase of the normalized emittance still affects these laser-plasma generated bunches due to the combination of high energy spread and divergence, and conventional magnet transport systems are not able to counteract this effect. We can conclude that to contrast the phenomenon it is necessary to act directly on the laser-plasma source, by improving the beam qualities, for example, with a laser-beam shaping.

\section{CONVENTIONAL STRATEGIES TO IMPROVE BEAM TRANSPORT}
\label{sec:cut}

If the total beam charge is of the order of magnitude as in Table \ref{tab1}, the current is about 100 kA, a factor between 100 and 1000 times higher than that required in photoinjectors for driving a FEL. The idea is then cutting the bunch, thus reducing the beam divergence and the energy spread at the cost of loosing some charge. The simplest way to do that is by using slits or apertures, placed in optimized points of the matching line, which cut transversely the beam.

A first rectangular aperture has been positioned at 60 cm from the laser-plasma interaction point, where the longitudinal magnetic field is zero. The aperture cuts the bunch horizontally at $\pm$0.4 mm and vertically at $\pm$0.3 mm. In this way we pass from the initial charge of 700 pC to about 230 pC. The initial current is reduced to about 40 kA, a still very high value, and the positive aspect is that the normalized emittance is drastically pulled down of a factor higher than 7.

\begin{figure}[!h]
\centerline{\includegraphics[width=7.5cm]{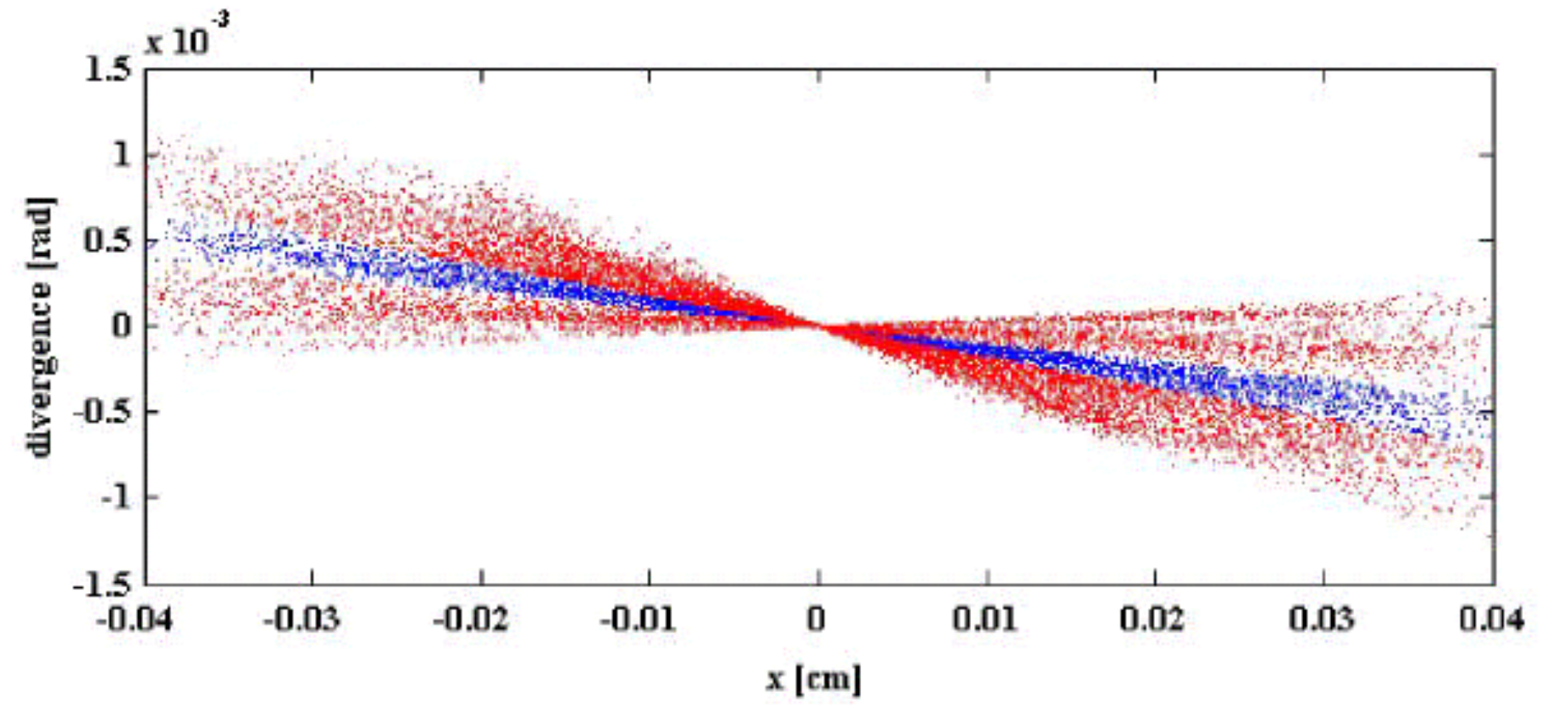}} 
\caption{(Color online) Transverse beam distribution obtained with TSTEP after the cut with the rectangular aperture: blue dots represent particles having energy spread less than 2\%, and red dots all the others.}
\label{fig10}
\end{figure}

With this cut alone, we cannot affect much the energy spread that remains almost unchanged, passing from 6.4\% to 6\%. It is interesting to note however that in this position there is a profitable correlation between energy and transverse phase space. In fact, in Fig. \ref{fig10} we show the transverse phase space after the cut with the blue dots representing all particles whose energy differs less than 2\% from the average and with the red dots all other particles.

\begin{figure}[!h]
\centerline{\includegraphics[width=7.5cm]{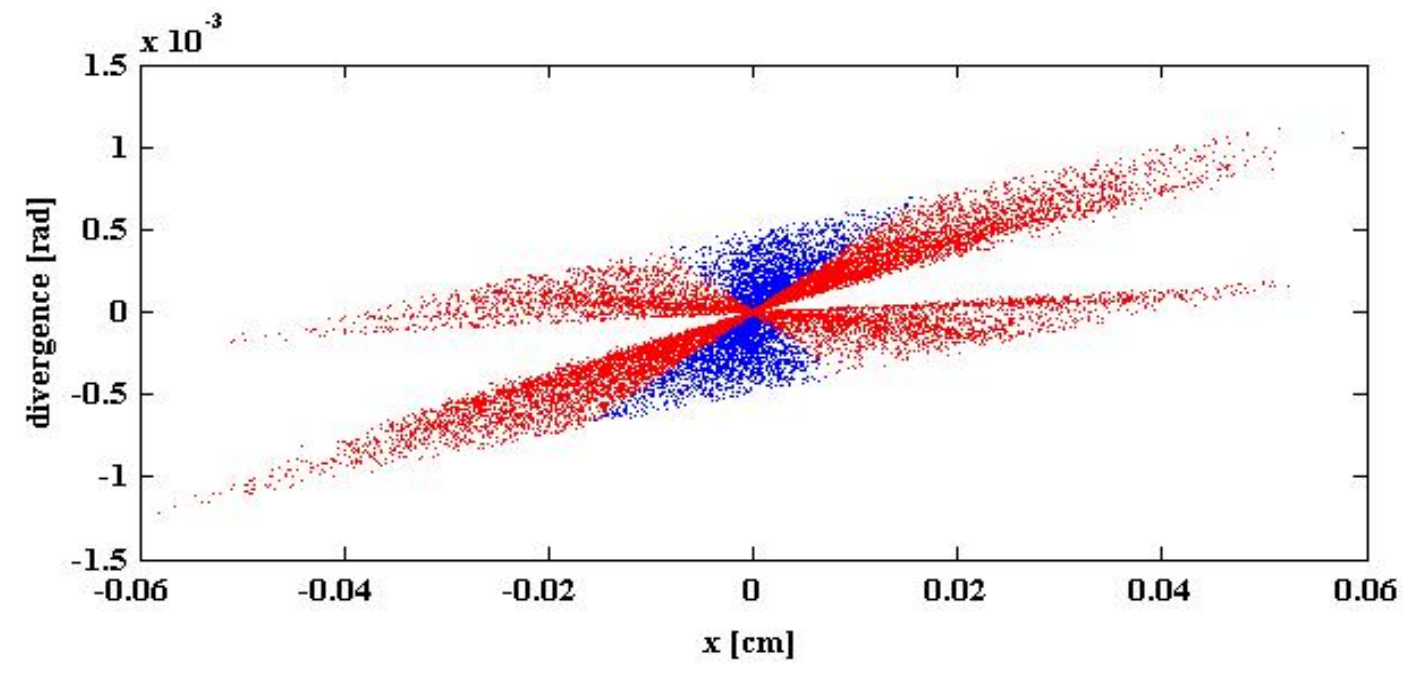}} 
\caption{(Color online) Transverse beam distribution obtained with TSTEP at 140 cm from the laser-plasma interaction point, with blue dots representing particles having energy spread less than 2\%, and red dots all the others.}
\label{fig11}
\end{figure}

Since in a drift all the particles having a divergence greater than zero move in the right direction of the figure, while particles with negative divergence move toward left, the bunch rotates, enlarging the phase space: a cut at the correct position then allows to reduce the energy spread. 
The same phase space after a drift of 80 cm (i.e. at 140 cm from the laser-plasma interaction point) is shown in Fig. \ref{fig11}. If we place here a rectangular aperture with the proper dimensions, we are able to cut particles having high energy spread.

This aperture has been chosen to be 80 $\mu$m large and 60 $\mu$m high, allowing us to obtain a bunch with an energy spread of 2.74\%, a charge of about 70 pC, and a peak current of 13.5 kA. Such a bunch has characteristics close to those of conventional accelerators and is not so hard to match to a FODO periodic line, since the Twiss beta-functions are about 20 cm. This matching line gives a normalized emittance of about 10 mm mrad, a factor 40 lower than that obtained without cuts. Moreover, the transverse cuts are quite easy to realize with simple rectangular apertures and, despite the particle loss, we end up with a quite high current.

\begin{figure}[!h]
\centerline{\includegraphics[width=6.8cm]{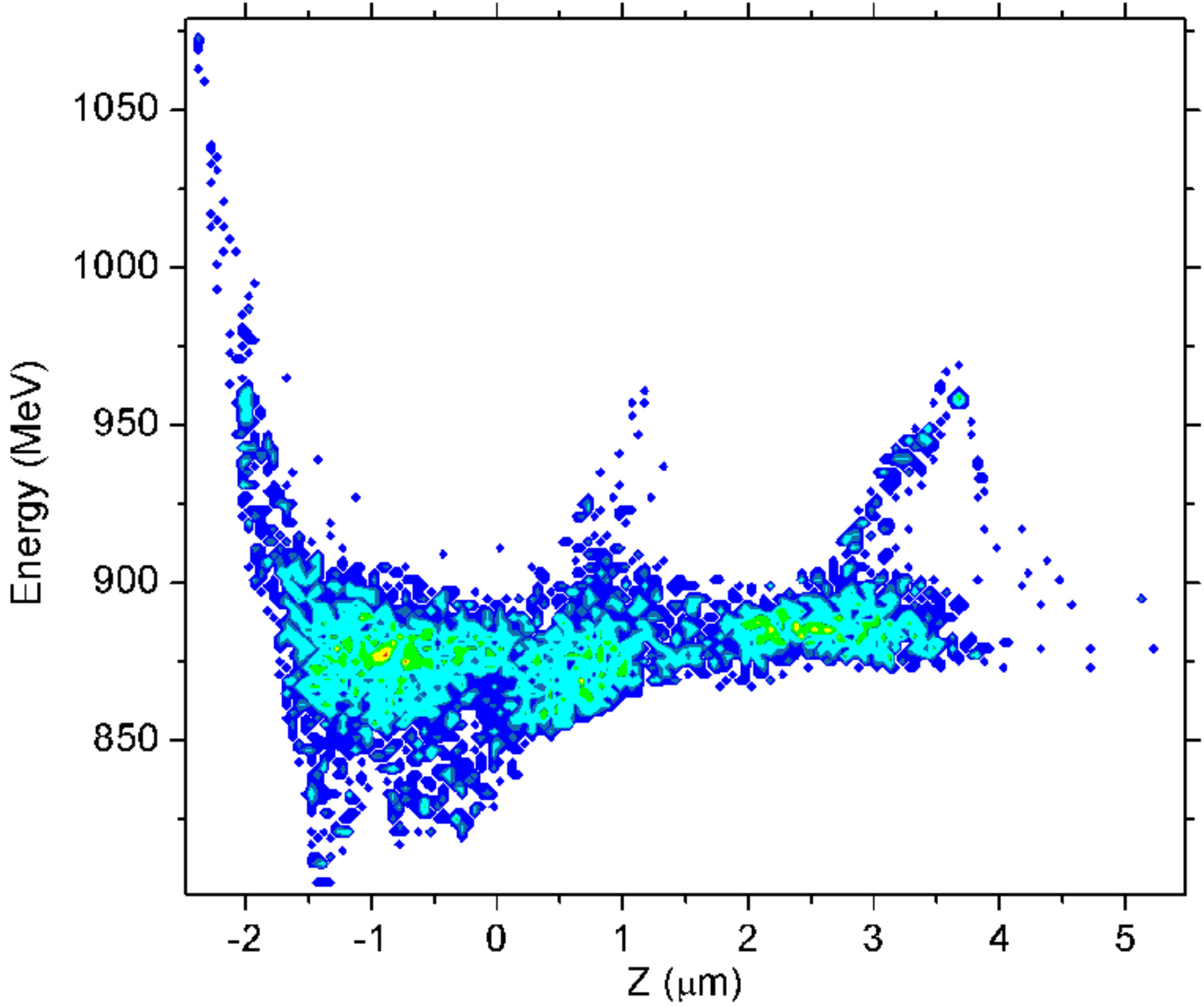}} 
\caption{(Color online) Longitudinal phase space particle distribution obtained with TSTEP after a second beam cut with transverse rectangular aperture.}
\label{fig12}
\end{figure}

However the energy spread can not be completely controlled. With transverse cuts, this quantity cannot be reduced to values less than 1\% - 2\% without a big loss of current. If we want to keep a quite high current, other mechanisms should be devised, e.g. a transport through a bending magnet or an isochronous system. In fact, Fig. \ref{fig12}, shows the longitudinal phase space particle distribution after the second aperture; particles with high energy ($\ge$ 1 GeV), far from the core of the bunch, give a high contribution to the energy spread, but a small one to the whole beam current.

Particles with such an energy distribution could not be cut by using simply transverse apertures, without first making the beam pass through a dispersion region, realized with bending magnets that produce energy dependent transverse trajectories. These can eventually be exploited to create a transverse beam-energy correlation that allows to properly cut the energy distribution. However the high beam qualities are ruined as the beam enters in a bending magnet, making again hard the control of the bunch characteristics. Moreover, given the high current, coherent synchrotron radiation can generate microbunching.

The procedure described above could fail to achieve the desired beam parameters on most shots because of intrinsic limits in the stability and repeatability  of self-injected beams. In particular, laser pointing instability and inhomogeneities in the plasma density produce a spread in the beam centroid direction, with respect to the design trajectory, that is in the order of some mrads. Since the second aperture on our transport line is smaller than the position spread expected at 140 cm, most shots would just be completely stopped. An alternative strategy could be tuning the transport line on a well definite energy, e.g. the neatly defined energy peak seen on the spectrum in Fig. \ref{fig4a}. We expect that the electrons with higher or lower energies, being unmatched, would be lost to the pipe walls, managing to produce a beam with low energy spread and not significantly affected by chromatic effects. However, this scheme can also fail, due to the same limits on stability and repeatability, since the neat peak energy inside the bunch, its central energy and the average energy of the whole bunch are all strongly affected by jitters typical of laser-plasma interaction.

Therefore, we can conclude that a great amount of work needs to be done on the production side of the plasma generated beams before reaching a quality compatible with conventional accelerator transport line. According to the discussion above, the plasma beam quality must be increased looking for strategies to reduce the energy spread, the beam divergence and beam centroid spread inside the plasma itself.

\section{APPLICATION TO Free Electron Lasers}
\label{sec:FEL}

\begin{figure}[!h]
\centerline{\includegraphics[width=7.5cm]{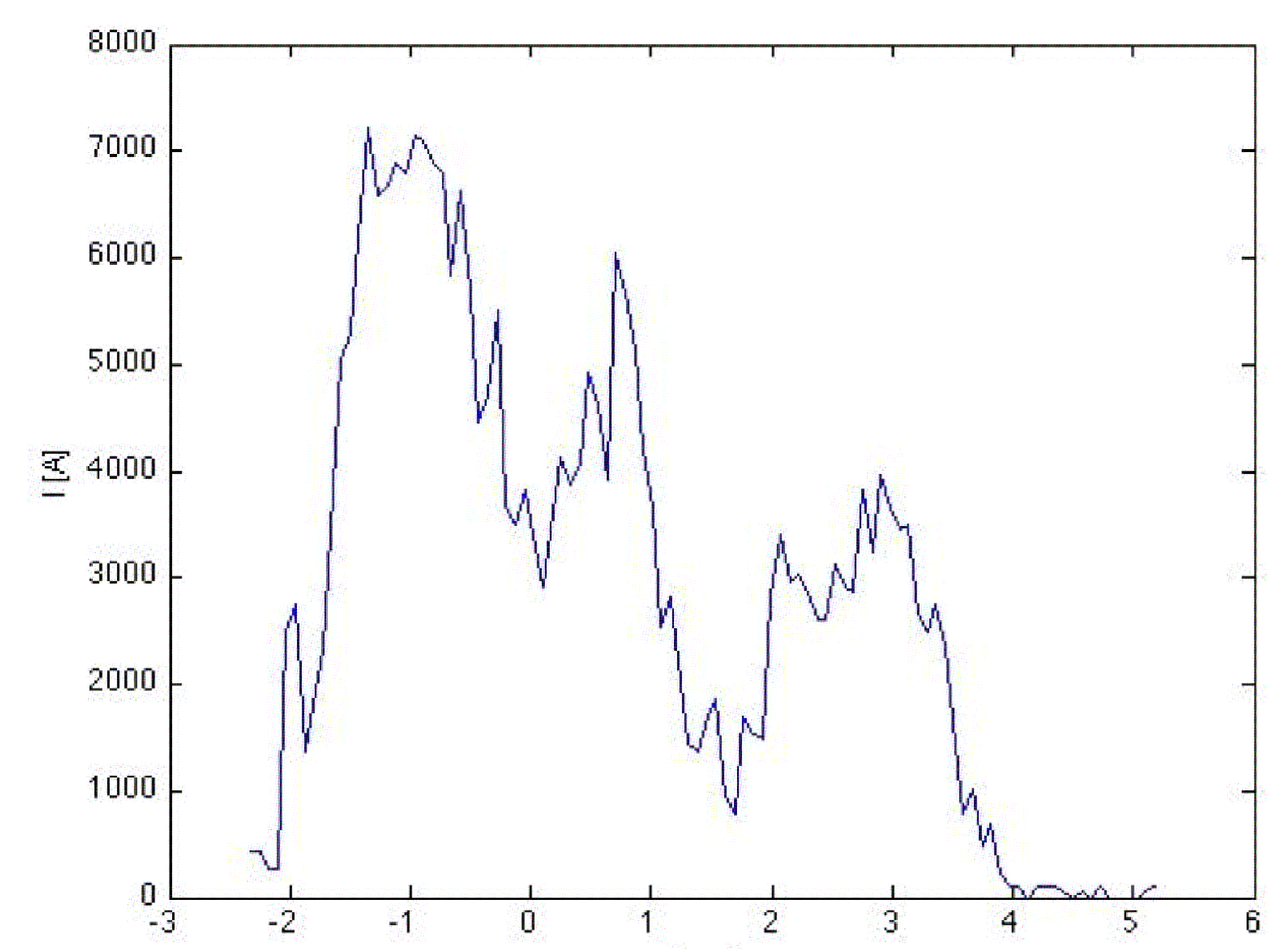}} 
\caption{(Color online) Beam current profile at the entrance of the undulator sections.}
\label{Fig:PeakCurr}
\end{figure}

In order to asses the potentiality of a laser-plasma generated beam, we have numerically studied the propagation of such a beam into undulators for the generation of hard X-rays; in the following, we deal with the beam after the matching discussed in Sec.~\ref{sec:cut}. The simulations of the output characteristics of the SASE FEL radiation,  such as temporal and spectral profiles, have been performed by means of the time dependent three-dimensional simulation code GENESIS 1.3 \cite{SvenGenesis}, accounting for slippage, diffraction, emittance and energy spread effects.
The SPARX Hard X-ray undulator beamline (undulator period and RMS parameter, $\lambda_u$=1.5 cm and $K_{RMS}$=0.907, respectively) \cite{SPARX-TDR} has been used to produce SASE FEL radiation at 4.6 nm at the fundamental wavelength, with electron beam average parameters reported in Table \ref{Tab:ebeamFEL} and beam current profile of Fig. \ref{Fig:PeakCurr} . 
%
\begin{table}[!h]
\centerline{
\begin{tabular}{||c|c||}
\hline\hline
Electron charge $Q$& 72 pC \\ 
\hline
Beam energy & 881 MeV \\
\hline
Energy spread (rms) & 2.74 \% \\
\hline
Bunch length (rms) $\sigma_{\tau}$ & 1.6 $\mu$m\\
\hline
Average current $Q/(\sqrt{2\pi} \sigma_{\tau})$ & 5.4 kA\\
\hline
$\sigma_x$, $\sigma_y$ & 33, 27 $\mu$m \\
\hline
$\sigma_{x'}$, $\sigma_{y'}$ & 0.2 mrad \\
\hline
$\varepsilon_{nx}$, $\varepsilon_{ny}$ & 12.8 mm mrad\\
\hline\hline
\end{tabular}
}
\caption{ Average beam parameters used for the SASE FEL simulation.}
\label{Tab:ebeamFEL}
\end{table}

\begin{figure}[!h]
\centerline{\includegraphics[width=7.5cm]{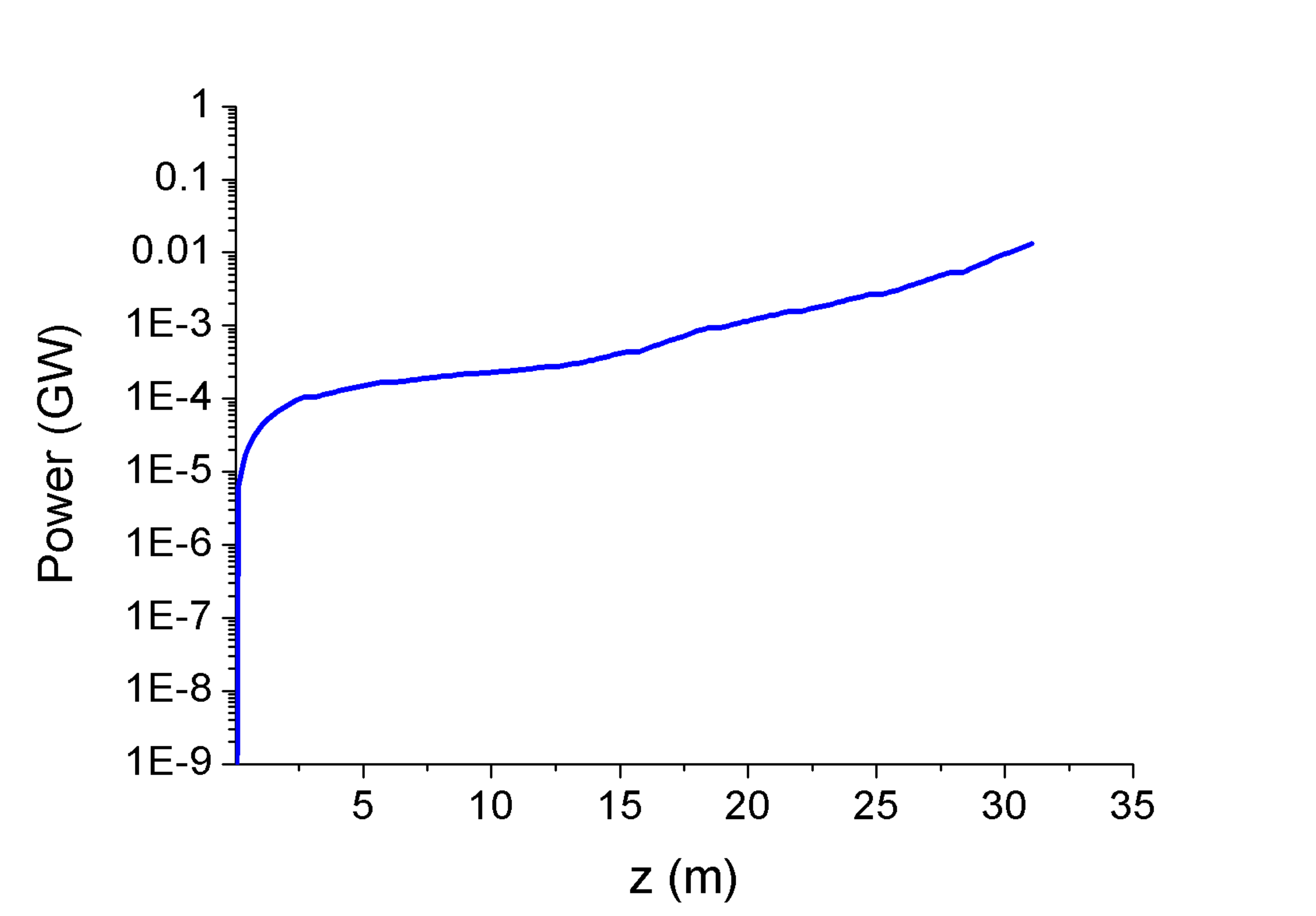}} 
\caption{(Color online) Maximum power in the radiation pulse along the undulator for the gaussian bunch case.}
\label{Fig:PowerTdepAve}
\end{figure}

\begin{figure}[!ht]
\centerline{\includegraphics[width=7.5cm]{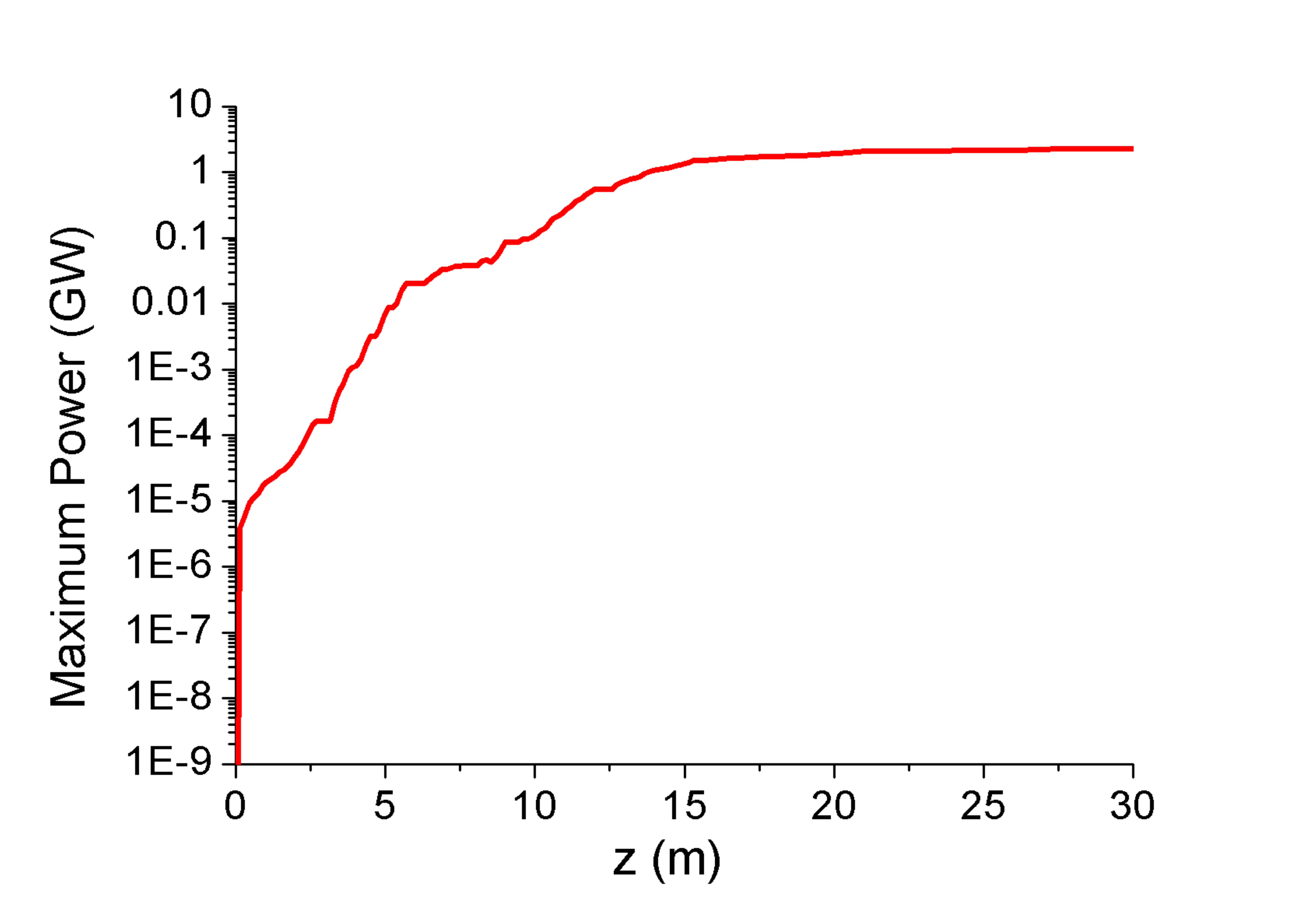}} 
\caption{(Color online) Maximum power in the radiation pulse along the undulator for the TSTEP-generated beam distribution.}
\label{Fig:PowerTdep}
\end{figure}

The saturation power along the undulator is shown in Fig. \ref{Fig:PowerTdepAve} for a bunch with Gaussian current distribution and with the parameters reported in Table \ref{Tab:ebeamFEL}. Figure~\ref{Fig:PowerTdep}, instead, is the saturation power when the actual bunch distribution obtained by a TSTEP simulation is injected into the undulator.
For the Gaussian bunch, the saturation length is much longer than the undulator length while, for actual one, the onset of saturation is evident at 15 m (being the 3D gain length about 0.7 m) because of the slices with higher current. In the simulation we noticed that the peak current decreases along the undulator because of particle losses due to both a non well optimized matching at the undulator entrance and large beam divergence. Therefore the current of the bunch actually contributing to the SASE FEL process is much smaller than the one shown in Fig. \ref{Fig:PeakCurr}; in fact, we would have expected an even smaller saturation length if the current distribution remained constant along the undulator and equal to Fig. \ref{Fig:PeakCurr}. The issue of such beams is that the excellent properties (e.g. brilliance) can not be exploited because of the difficulties in matching (due to the big energy spread and divergence). 

\begin{figure}[!h]
\centerline{\includegraphics[width=7.5cm]{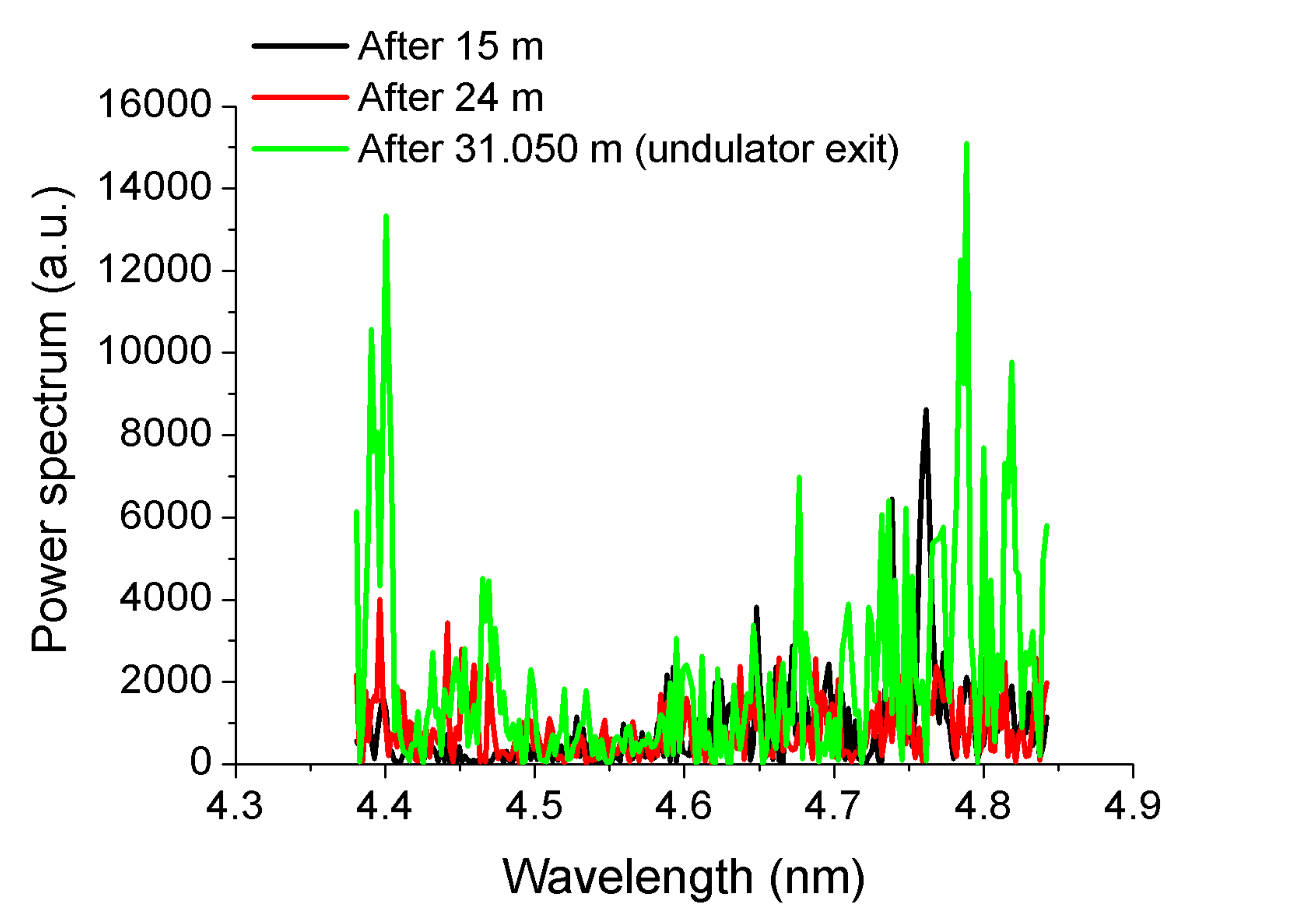}} 
\caption{(Color online) Spectral structure for 4.6 nm radiation from the SPARX Hard X-ray undulator beamline at saturation (31 m) in case of the external distribution from TSTEP.}
\label{Fig:Spectrum31m}
\end{figure}

The spectral analysis for the 4.6 nm radiation is shown in Fig. \ref{Fig:Spectrum31m} at the saturation (15 m, black curve), after 24 m (red curve) and at the end of the undulator (green curve). The emission is far from being monochromatic and it is centered around the fundamental wavelength, with a 3\% radiation bandwidth spread. 

Our simulations have demonstrated that in spite of large emittance and huge energy spread, thanks to the high peak current, SASE FEL saturation occurs even in moderate undulator lengths (despite the significant portion of charge lost in the propagation in the undulator). However, the radiation spectrum has not the shape typical of conventional SASE FELs (spike radiation) because the energy spread is too high (even higher than the FEL $\rho$ parameter). 

\section{CONCLUSIONS}

In this paper, we have shown that laser-plasma generated beams possess peculiar characteristics with respect to the conventional accelerator beams: for example in a drift space, the normalized projected emittance increases due to the high energy spread and divergence. We have demonstrated that capture of such beams with conventional accelerator devices is in principle feasible:  one can use solenoids and rectangular apertures to properly shape the beam, even despite some beam charge losses, and devise an isochronous beam line to further reduce the energy spread. Applications, such as FELs (i.e. requiring good slice beam properties), can profit of the fact that slice emittance may not be diluted too much since slice energy spread can be quite low for such beams.
Anyway we believe that the qualities of such beams (in terms of energy spread and beam divergence) must be improved first inside the plasma itself (e.g. properly shaping the plasma channel, control over injection, etc...) before allowing the use in conventional accelerator lines.

\section*{Acknowledgements}

The authors express their appreciation to the SPARC group for many clarifying discussions. We also acknowledge the support of ELI-PP.



\begin{thebibliography}{99}

\bibitem{uno} \emph{The Future of Accelerator Physics}, edited by T. Tajima, AIP, NY, (1996).

\bibitem{Esa09} E. Esarey et al., \emph{Physics of laser-driven plasma-based electron accelerators}, Rev. Mod. Phys. \textbf{81}, Issue 3, p. 1229 (2009).

\bibitem{Ged04} C.G.R. Geddes et al., \emph{High-quality electron beams from a laser wakefield accelerator using plasma-channel guiding}, Nature {\bf 431}, p. 538 (2004).


\bibitem{Lee06} W.P. Leemans et al., \emph{GeV electron beams from a centimetre-scale accelerator}, Nature Physics {\bf 2}, p. 696 (2006).

\bibitem{Gor05} S. Gordienko and A. Pukhov, \emph{Scalings for ultrarelativistic laser plasmas and quasimonoenergetic electrons}, Phys. Plas. {\bf 12}, 043109 (2005).

\bibitem{Lu07} W. Lu et al., \emph{Generating multi-GeV electron bunches using single stage laser wakefield acceleration in a 3D nonlinear regime}, Phys. Rev. ST - Acc. Beams - {\bf 10}, 061301 (2007).

\bibitem{Bru10}E. Brunetti et al., \emph{Low Emittance, High Brilliance Relativistic Electron Beams from a Laser-Plasma Accelerator}, Phys. Rev. Lett. {\bf 105}, 215007 (2010).

\bibitem{Sea10} C.M.S. Sears et al., \emph{Emittance and divergence of laser wakefield accelerated electrons}, Phys. Rev. ST - Acc. Beams - {\bf 13}, 092803 (2010).

\bibitem{Rec09} C. Rechatin et al., \emph{Controlling the Phase-Space Volume of Injected Electrons in a Laser-Plasma Accelerator}, Phys. Rev. Lett. {\bf 102}, 164801 (2009).

\bibitem{Gon11} A.J. Gonsalves et al., \emph{Tunable laser plasma accelerator based on longitudinal density tailoring}, Nature Physics {\bf 7}, 862 (2011).

\bibitem{r1} M. Reiser,\emph{ Theory and Design of Charged Particle Beams}, second edition, WILEY-VCH Verlag GmbH \& Co. KGaA, Weinheim, p. 105, Eq.~(3.162).

\bibitem{flame} See the website \texttt{http://ilil.ipcf.cnr.it/flame}.

\bibitem{Ben09} C. Benedetti, \emph{Simulation of particle acceleration in the PLASMONX project}, Proc. of the 2nd International Conference on ULTRA-INTENSE LASER INTERACTION SCIENCE, Frascati, Italy, May 24-29, 2009, AIP Conf. Proc. 1209, 11-14 (2010).

\bibitem{aladyn} C. Benedetti, et al., \emph{ALaDyn: A High-Accuracy PIC Code for the Maxwell-Vlasov Equations},
IEEE - Trans on Plasma Science {\bf 36}, N. 4, 1790(2008). C. Benedetti et al., \emph{PIC simulations of the production of high-quality electron beams via laser-plasma interaction}, Nuc. Inst. Meth. in Phys. Res. A, 608, S94-S98 (2009)



\bibitem{tstep} L. M. Young, priv. comm.

\bibitem{parmela} L. M. Young, \emph{PARMELA}, Los Alamos National Laboratory report LA-UR-96-1835.

\bibitem{emit} \emph{Advances in Solid-State Lasers: Development and Applications}, Edited by Mikhail Grishin, 600 (2010).

\bibitem{sparc} M. Ferrario, et al., \emph{Recent results of the SPARC project}, Proceedings of FEL08, Gyeongju, Korea, 359 (2008).

\bibitem{r2} M. Reiser,{\em ibid.}, par. 3.1.

\bibitem{Cou72} E.D. Courant, \emph{Impossibility of  achromatic focusing with magnetic quadrupoles and solenoids}, Part. Acc. {\bf 2}, p. 117 (1972).

\bibitem{Tom_ip} P. Tomassini et al., paper in preparation.

\bibitem{trace} K. R. Crandall, D. P. Rusthoi, \emph{TRACE 3-D documentation}, LA-UR-97-886, (1997).

\bibitem{pmq} S. Becker, et al., \emph{Characterization and tuning of ultrahigh gradient permanent magnet quadrupoles}, Phys. Rev. ST - Acc. Beams - {\bf 12}, 102801 (2009).

\bibitem{sparc1} J.B. Rosenzweig, et al., \emph{RF and Magnetic Measurements on the SPARC Photoinjector and Solenoid at UCLA}, PAC05 Proc., Knoxville, Tennessee, 2626 (2005).

\bibitem{lcls} J. Schmerge, \emph{LCLS Gun Solenoid Design Considerations},  LCLS-TN-05-14, (2005).

\bibitem{50t} S. A. Kahn, et al., \emph{HTS development for 30-50 T final muon cooling solenoids}, PAC09 Proc., Vancouver, BC, Canada (2009).

\bibitem{SvenGenesis} S. Reiche, \emph{Genesis 1.3 user manual}, (2004).

\bibitem{SPARX-TDR} \emph{SPARX Technical Design Report}, July 2009.

\end{thebibliography}
\end{document}